\documentclass[journal=ancat3,manuscript=communication,layout=twocolum]{achemso}

\usepackage[version=3]{mhchem} 
\usepackage{amsmath, amssymb}

\newcommand{\R}{R^\text{tot}}
\newcommand{\Rc}{R_\text{bg}}
\newcommand{\erf}{\operatorname{erf}}
\newcommand{\single}{\mathrm{s}}
\newcommand{\double}{\mathrm{d}}

\newcommand{\erfi}{\operatorname{erfi}}
\newcommand{\Vbar}{V_\text{barrier}}
\newcommand{\Vch}{V_\text{channel}}
\newcommand{\eL}{e^{-\beta F L_x}}
\newcommand{\ea}{e^{-\beta F a}}
\newcommand{\eaL}{e^{-\beta F (L_x-a)}}
\newcommand{\derf}{\Delta E}
\newcommand{\derfi}{\Delta I}

\renewcommand{\d}[2][]{\!\!\mathrm{d}^{#1}#2\;}

\newcommand{\hyp}[1]{\,{}_2F_2\left(1,1;\frac{3}{2},2; #1\right)}

  \author{Urs Zimmermann}
  \email{urs.zimmermann@uni-duesseldorf.de}
  \affiliation{Institut f\"ur Theoretische Physik II: Weiche Materie, Heinrich-Heine-Universit\"at D\"usseldorf, D-40225 D\"usseldorf, Germany}
  \author{Hartmut L\"owen}
  \affiliation{Institut f\"ur Theoretische Physik II: Weiche Materie, Heinrich-Heine-Universit\"at D\"usseldorf, D-40225 D\"usseldorf, Germany}
  \author{Christian Kreuter}
  \affiliation{Fachbereich Physik, Universit\"at Konstanz, D-78457 Konstanz, Germany}
  \author{Artur Erbe}
  \affiliation{Institut f\"ur Ionenstrahlphysik und Materialforschung, Helmholtz-Zentrum Dresden-Rossendorf, D-01328 Dresden, Germany}
  \author{Paul Leiderer}
  \affiliation{Fachbereich Physik, Universit\"at Konstanz, D-78457 Konstanz, Germany}
  \author{Frank Smallenburg}
  \affiliation{Institut f\"ur Theoretische Physik II: Weiche Materie, Heinrich-Heine-Universit\"at D\"usseldorf, D-40225 D\"usseldorf, Germany}
  \alsoaffiliation{Universit\'e Paris-Saclay, CNRS, Laboratoire de Physique des Solides, 91405 Orsay, France}

\title{Climbing two hills is faster than one: collective barrier-crossing by colloids driven through a microchannel}

\begin{document}







\begin{abstract}
 Ohm's law is one of the most central transport rules stating that the total resistance of sequential single resistances is additive. While this rule is most commonly applied to electronic circuits, it also applies to other transport phenomena such as the flow of colloids or nanoparticles through channels containing multiple obstacles, as long as these obstacles are sufficiently far apart. Here we explore the breakdown of Ohm's law for fluids of repulsive colloids driven over two energetic barriers in a microchannel, using real-space microscopy experiments, particle-resolved simulations, and dynamical density functional theory. If the barrier separation is comparable to the particle correlation length, the resistance is highly non-additive, such that the resistance added by the second barrier can be significantly higher or lower than that of the first. Surprisingly, in some cases the second barrier can even add a {\it negative} resistance, 
such that two identical barriers are easier to cross than a single one. We explain this counterintuitive observation in terms of the structuring of particles trapped between the barriers.
\end{abstract}

\section{Introduction}

One of the basic characteristics of any transport situation is the resistance, commonly known from electric circuits, which is in general defined as the ratio of the transport flux and the
driving force, typically in the linear-response regime of small drives. 
For both electric circuits and classical transport, Ohm's law states that when resistors are put in series, their resistances simply add up.
However, this macroscopic law is expected to break down on the microscopic scale, in particular 
when the distance between the two obstacles approaches the correlation length of the transported particles.
 
Knowing and controlling flow resistance is of particular importance when tuning the transport of solutes through channels. This type of transport is the basic situation in microfluidics \cite{micro}, where the transported objects are typically micron-sized colloidal solutes, such that 
thermal fluctuations play a significant role \cite{Marchesoni_Haenggi_RMP}. 
Similar transport scenarios include the collective migration of bacteria through channels \cite{PhysRevLett.101.018102,Wensink_PNAS_2012,Wensink_HL_PRE_2008}, the transport of nanoparticles through porous media \cite{dunphy2006influence, chowdhury2011mechanisms}, and the transport of ions through membranes via nanopores \cite{dzubiella2005electric}. On the macroscopic scale, the flow of e.g. cars or pedastrians \cite{Helbing_review} or animals \cite{animals} through crowded environments can lead to similar physics.
Obstacles in such channels naturally inhibit the overall steady-state rate at which the particles 
are able to traverse the channel, providing an effective resistance to the flow.
In channels with multiple obstacles, we expect Ohmic (i.e.\ additive) behavior of the corresponding resistances when the separation between the obstacles
is large, and a breakdown of Ohm's law for smaller distances. The crossover between these regimes is determined by the correlation length in the system, i.e. the length scale associated with local structure in the fluid of transported particles. Detailed knowledge of these non-additive effects is of vital importance for the design of efficient microfluidic devices, as well as for our broader understanding of constricted flow phenomena.
 
Here, we explore the additivity of resistances in mesoscopic colloidal suspensions driven through a microchannel \cite{Microchannels_I}. 
First, as a proof of concept, we perform an experiment on repulsive colloidal particles confined to microchannels containing two step-like barriers on the substrate, and measure the current through the channels as a function of the strength of the gravitational driving force. Our results show that step-like barriers in a microchannel can indeed be interpreted as resistors. We then further explore this concept using Brownian dynamics simulations and dynamical density functional theory and map out the interplay between the two barriers by varying their height and separation. We find strong deviations from additivity for the resistance of two barriers when the separation between the two obstacles is comparable to the correlation length of the system, which is on the order of several interparticle spacings. Amazingly, if the barrier separation is comparable to the interaction range, the resistance contributed by the second barrier can even be {\it negative}
such that climbing two hills is faster than one. We explain this counterintuitive effect of negative resistance 
via the long-ranged particle interactions and the ordering of the particles trapped between the two barriers. When these particles are disordered, they exhibit spontaneous fluctuations which modulate their interactions with particles crossing the barriers, significantly enhancing barrier crossing rates \cite{stein1989mean,pechukas1994rates}. This surprising phenomenon provides a route for tuning and enhancing particle flow over an obstacle by the inclusion of additional barriers, reminiscent of the use of geometric obstacles to assist e.g. the flow of panicked crowds \cite{shiwakoti2013enhancing}.

\section{Results}

\subsection{Experiment}

\begin{figure*}
 \hspace{2.3cm} {\large d)}\\[-0.5cm]
    \begin{tabular}{ccc}
       \includegraphics[height=0.5\linewidth]{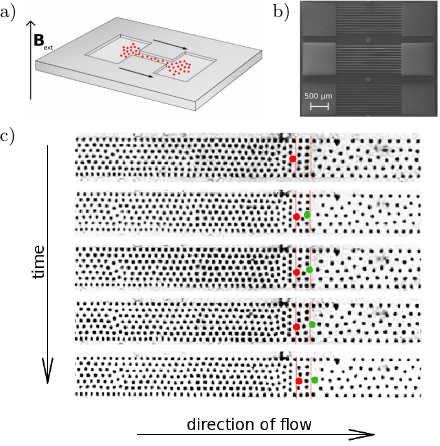} & \,\,\,&
       \includegraphics[height=0.5\linewidth]{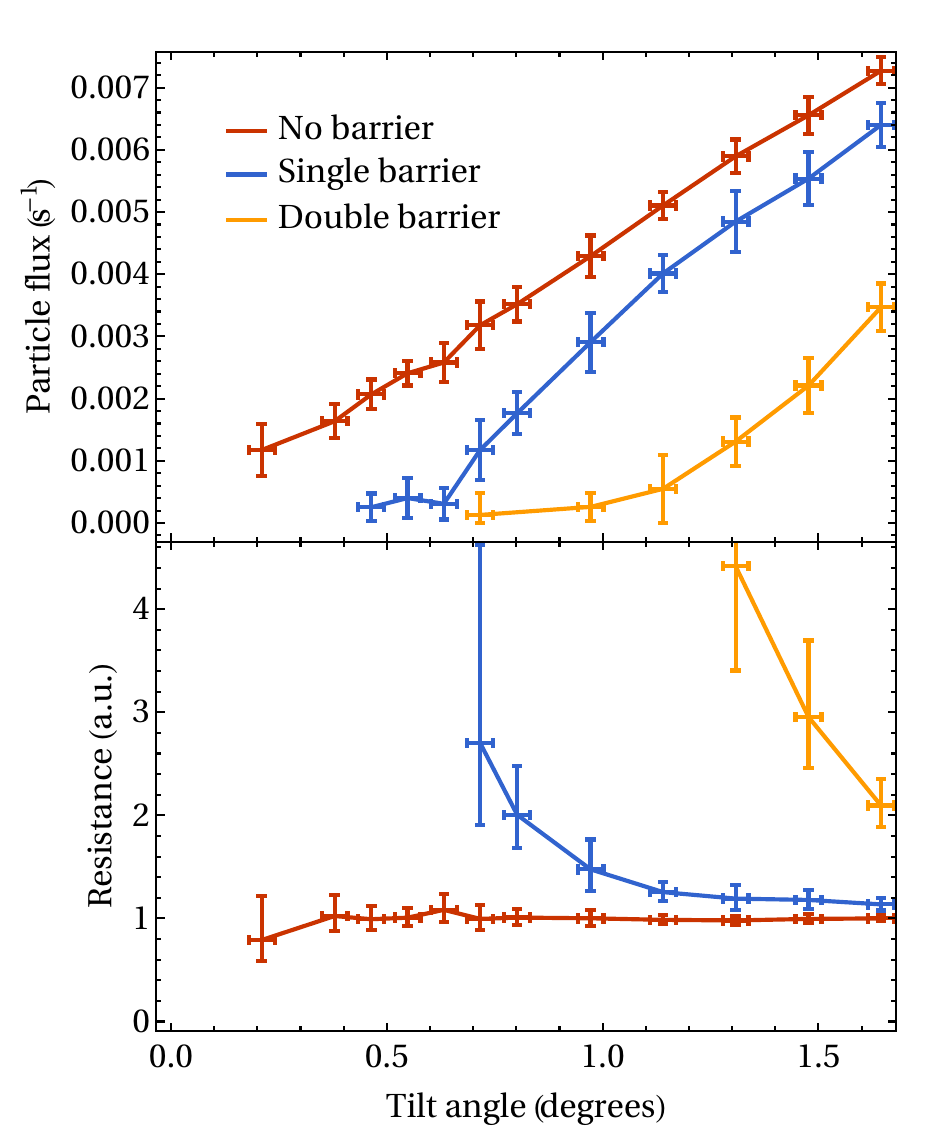}
    \end{tabular}
 \caption{{\bf Experimental proof-of-concept.} a) Schematic setup of the experiment: two particle reservoirs are connected by a microfluidic channel
through which particles are flowing due to gravity. b) Top view of the experimental system.
c) Snapshots of a two barrier system at different times. The position of the barriers is indicated by a red vertical line. Two particles are highlighted in red and green. 
d) Particle flux (top) and the effective resistance of the second barrier (bottom) as a function of the tilt angle in a system with no barriers (red), single barrier (blue) and double barriers (orange). The initial density was $\rho_0 = (7.23 \pm 0.5)\times 10^{-3} \mu\text{m}^{-2}$, the external field strength was $0.6$ mT, and the separation between the two barriers was $30 \mu$m.
}
 \label{fig:experiment}
\end{figure*}

We measure the particle current in the channel as a function of the gravitational driving force, controlled by the tilt angle of the setup, for channels with zero, one, and two barriers. In the absence of barriers, the current shows the expected linear dependence on the driving force, shown by the red line in Fig.\ \ref{fig:experiment}d. For a single barrier (blue line in Fig.\ \ref{fig:experiment}d), we observe a crossover from a zero-flow regime at small driving forces (where the driving force is too weak to push particles across the barrier) to an approximately linear regime for large driving forces \cite{Microchannels_I}. Hence, the barrier provides a resistance to the flow, which reduces the particle current.  
Adding a second barrier to the channel clearly results in a further decrease of the current, as one would expect (orange line in Fig. \ref{fig:experiment}d). In order to examine the possibility of non-additive resistance, we also plot in Fig. \ref{fig:experiment}d the total resistance of the channel, defined as the current divided by the driving force. Here, we only consider tilt angles where the channels do not get fully blocked. For an empty channel, we find a well-defined constant resistance, consistent with the linear behavior of the current as a function of the tilt angle. The single barrier increases this background resistance. For sufficiently large driving forces, this increase is essentially constant, indicating that we can indeed interpret it as a simple additional resistor added to the channel. For low driving forces, the effective resistance added by the barrier is significantly higher, which we attribute to intermittent blocking of the channel: in this regime, the driving force is not always capable of pushing the particles across the barrier, and thermal fluctuations likely play a role in enabling the flow. Finally, adding a second barrier adds another contribution to the total resistance. Interestingly, this additional resistance is not simply equal to the resistance of the first barrier, even though we are in the regime where the particles flow through the channel without blockage. In particular, the total flux in the two-barrier system at the highest tilt angle is on the same order as the flux in the single-barrier case when its resistance has reached its plateau value (see Fig. \ref{fig:experiment}d). Hence, we conclude that the two resistors interact non-additively in this case, indicating a breakdown of Ohm's law. As the barriers in this experiment are separated only by a distance of approximately 2.5 times the typical interparticle distance, which is shorter than the correlation length in the fluid, this breakdown could be the result of microscopic structuring of the fluid between the two barriers. Indeed, as shown by the snapshots in Fig. \ref{fig:experiment}c, we consistently find two layers of particles in between the barriers. To explore this concept of non-additivity further, we now turn to a numerical treatment of the problem, where we can more readily explore a wide range of conditions.

\subsection{Theory and Simulation}
We make use of overdamped Brownian dynamics simulations and dynamical density functional theory (DDFT) calculations. We consider a two-dimensional system with periodic boundary conditions along the channel ($x$-direction), containing $N$ particles interacting via a dipolar repulsion
\begin{equation}
 \beta V_\text{int}(r) = \Gamma \left(\frac{a}{r}\right)^3,
\end{equation}
where $\beta = 1 / k_B T$ with $k_B$ Boltzmann's constant and $T$ the temperature, $\Gamma$ is the dimensionless interaction strength, and $a=\rho_0^{-1/2}$ sets the length scale of a typical interparticle spacing of a given mean number density $\rho_0$. The particles additionally experience a constant driving force $F \hat{\mathbf{x}}$ pushing the particles along the channel. 

\begin{figure}[tbh]
\centering
\includegraphics[width=0.8\linewidth]{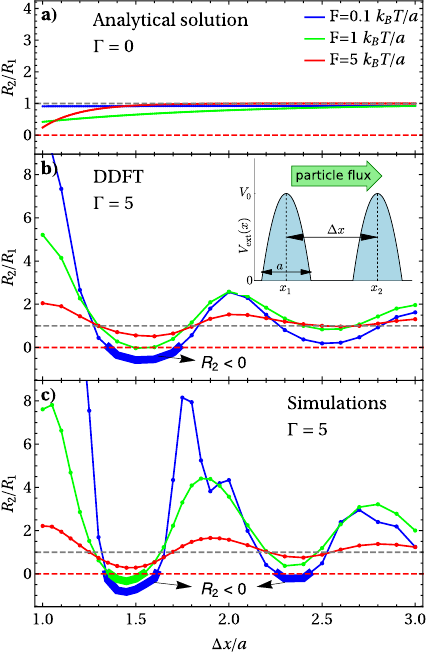}
\caption{{\bf Effective resistance of a second barrier.} Effective resistance $R_2$ of the second barrier relative to the resistance $R_1$ of the first barrier, as a function of the barrier spacing $\Delta x$, at different driving forces. The dashed lines highlight special values of $R_2$: the gray line shows Ohmic additivity and the red line marks the onset of negative effective resistance. Results are shown for analytical theory at $\Gamma=0$ (a), DDFT at $\Gamma = 5$ (b), and simulations at $\Gamma = 5$ (c). A sketch of the barrier configuration is shown in inset b.}
\label{fig:Rplots}
\end{figure}

The confining channel and barriers are modeled as an external potential $V_\text{ext}(x,y) = V_\text{channel}(y) + V_{\text{barrier}}(x)$. The first term here is a steep repulsive wall potential confining the particles in one direction. $V_\text{barrier}$ represents one or two parabola-shaped potential barriers with width $a$ and height $V_0 = 10 \, k_\text{B}T$, see Fig.\ \ref{fig:Rplots}b inset and Methods.
We choose the channel width $L_y = 4.65 a$, and the channel length $L_x$ such that the total number density $\rho_0 = N / (L_x L_y) = 1/a^{2}$ for a given particle number $N$.

In our DDFT calculations \cite{DDFT_Tarazona,DDFT_Evans}, we choose the Ramakrishnan--Yussouff functional \cite{RY_functional} to model interacting particles in a fluid state ($\Gamma = 5$). In addition to DDFT, we perform Brownian Dynamics simulations of particles experiencing the same potentials and external driving force. As a reference we provide an analytical solution for non-interacting particles ($\Gamma=0$). See Methods for details.

Using both DDFT and simulations, we explore the relation between the total steady-state particle current $J$ along the channel, the driving force $F$ on the particles, and the distance $\Delta x$ between the two barriers. The ratio of the driving force and current characterizes the total resistance of the system, $\R = \frac{F}{J}$.
In a channel without barriers, the particles trivially adopt the average drift current $J_0 = F \rho_0 L_y \xi^{-1}$, where $\xi$ is the friction coefficient of the background solvent, leading to an inherent background resistance $\Rc = \xi / (L_y\rho_0)$.
In a single-barrier system, the resistance $R_1$ added by the barrier can be extracted from the total resistance $\R_\single = \Rc + R_1$ by measuring the single-barrier current $J_\single$:
\begin{align}
 R_1 = \R_\single - \Rc = F\left(\frac{1}{J_\single} - \frac{1}{J_0}\right).
\end{align}
Similarly, in a double-barrier system (with current $J_\double$), the total resistance is $\R_\double = \Rc + R_1 + R_2$, and the effective resistance of the second barrier $R_2$ can be written as
\begin{align}
 R_2 = F\left(\frac{1}{J_\double} - \frac{1}{J_\single}\right).
\end{align}
In the case of additivity, the resistance $R_2$ of the second barrier will be equal to $R_1$ (the resistance of the first barrier), while deviations from this rule will indicate non-additivity.

In Fig.\ \ref{fig:Rplots}, we plot $R_2/R_1$ for a range of barrier separations $\Delta x$ at different driving forces $F$, as obtained from analytical theory (see Methods) (a), DDFT calculations (b), and computer simulations (c). For non-interacting particles $R_2$ is lowest when the two barriers are touching ($\Delta x = a$) and converges exponentially to $R_1$ for larger distances. In contrast, for interacting particles and for all investigated $F$, the resistance of the second barrier is highest at $\Delta x = a$. At this separation the resistance added by the second barrier can be many times higher than $R_1$, signaling strong non-additivity. More interestingly, for slightly larger separations ($\Delta x \simeq 1.5 a$), $R_2$ becomes smaller than $R_1$, and even negative for sufficiently weak driving forces. In this regime, the addition of the second barrier {\em reduces} the overall resistance in the channel. At larger $\Delta x$, $R_2$ shows decaying oscillations,  converging towards the additive case ($R_2 = R_1$), as expected at sufficiently large distances.

\begin{figure}
 \includegraphics[width=0.95\linewidth]{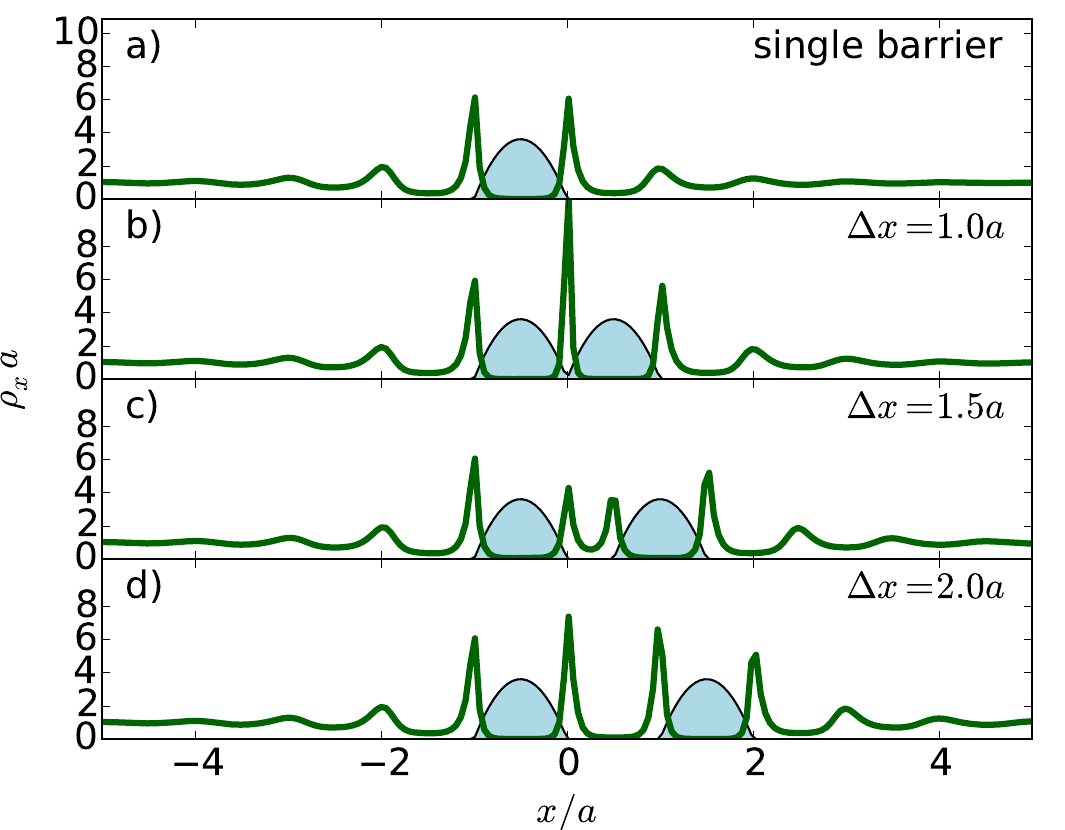}
 \caption{{\bf Density profiles near the barriers.} Local density profiles as a function of distance along the channel at the same interaction strength ($\Gamma = 5$) and driving force $F = 0.1 k_B T / a$, as obtained via DDFT. From top to bottom, we show a system with a single barrier, and systems with two barriers at separations $\Delta x / a = 1.0, 1.5$, and $2.0$.  }
 \label{fig:densprofs}
\end{figure}

We can understand this observation by considering the interactions between the particles. Since these are dipolar in nature, they are sufficiently long-ranged to span across the barrier. Hence, a particle on top of the barrier experiences forces from particles between the two barriers, which depend on the density and structuring of those particles. In Fig.\ \ref{fig:densprofs} we plot the density profile of the particles  $\rho_x(x)$, projected onto the long axis of the channel, for various barrier separations $\Delta x$, as well as for a single barrier. In the single-barrier case, we always observe a high density peak in front of the barrier, and a slightly lower peak just after the barrier (see Fig.\ \ref{fig:densprofs}a). In the two-barrier cases, the additional peaks in between the two barriers vary in height based on $\Delta x$. For very small separations (Fig.\ \ref{fig:densprofs}b), where the resistance of the second barrier is high ($R_2 > R_1$), we find a single sharp density peak between the barriers, which is significantly higher than the peak observed after a single barrier. Here, particles between the barriers are arranged in a single line with little room for fluctuations, and hence provide a strong and relatively constant force on particles crossing the first barrier, pushing them back. In the regime where $R_2 < R_1$ (Fig.\ \ref{fig:densprofs}c), we instead see two much lower peaks, indicating a structure with two layers and significantly larger fluctuations. These larger fluctuations not only provide space for particles entering via the first barrier, but also modulate the force exerted on particles crossing the barriers, resulting in a fluctuating effective barrier height. For weak driving forces, barrier crossings are rare events, whose rate depends exponentially on the barrier height. Fluctuations in barrier height are known to lead to significantly higher crossing rates \cite{stein1989mean,pechukas1994rates} and hence higher currents. Finally, for larger separations, where $R_2 > R_1$ again, we observe two higher peaks, indicating a more structured pair of layers between the barriers.

\begin{figure}
 \includegraphics[width=0.95\linewidth]{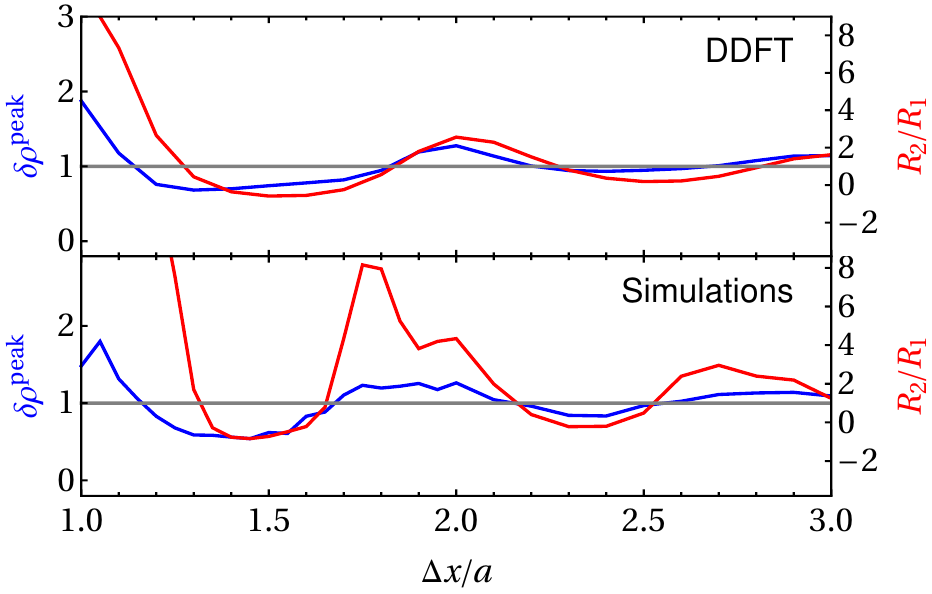}
 \caption{{\bf Ordering between the barriers.} Height of the first density peak (in blue) after the first barrier (normalized by the height of the peak after a single barrier) as a function of barrier separation at fixed driving force $F = 0.1 k_BT / a$ and interaction strength $\Gamma = 5$, as obtained from DDFT (top) and simulations (bottom). For comparison, we also plot the effective resistance of the second barrier in red.}
 \label{fig:peaks}
\end{figure}

We confirm this intuitive picture by plotting in Fig.\ \ref{fig:peaks} the relative height of the first peak after the first barrier $\delta \rho^\mathrm{peak} = \rho_\double^\mathrm{peak} / \rho_\single^\mathrm{peak}$, where $\rho_\single^\mathrm{peak}$ is the height of the first peak after a single barrier, and $\rho_\double^\mathrm{peak}$ is the height of the first peak after the first of two barriers. When plotted as a function of $\Delta x$, the peak height (blue in Fig.\ \ref{fig:peaks}) indeed strongly correlates with the particle current (red) in both the DDFT framework and the simulations. In our particle-resolved simulations, the additional fluctuations of the particles in between the two barriers are clearly visible. Moreover, examining simulation trajectories demonstrates that for most barrier separations, whenever a particle crosses the first barrier, the sudden increase in density between the barriers typically leads to the rapid expulsion of a particle over the second barrier. This observation confirms that the first of the two barriers can indeed be considered as the main bottleneck for the overall flow process. However, for $\Delta x \lesssim 1.3$, the bottleneck is instead the crossing of the {\it second} barrier. Here, particles form a single narrow layer between the two barriers, which inhibits the possibility of collectively pushing a particle across the second barrier. This may explain the reduced correlation between $\delta \rho_\mathrm{peak}$ and $R_2 / R_1$ for small $\Delta x$ in Fig.\ \ref{fig:peaks}.

\section{Discussion}

We have explored the effect of sequential potential energy barriers on the flow of colloidal particles driven through microchannels. As our experiment shows, two barriers close together can result in drastically higher resistance than twice the resistance of a single barrier. Moreover, via a detailed investigation of this non-additivity using both simulations and dynamical density functional theory, we discover that depending on the barrier spacing, the second barrier can add an effective resistance that is higher than the resistance of a single barrier, lower, or even negative. In the negative regime, the presence of the second barrier helps particles cross the first barrier, contrary to what intuition would suggest. We show that this enhanced barrier-crossing rate can be attributed to the structuring of the layer of particles in between the two barriers: weaker structuring (evidenced by lower peaks in the density profile) increase the current. A vital component for this phenomenon is the requirement that particles on top of the barriers can still interact with the particles aggregated just before and after that barrier, necessitating sufficiently long-ranged interactions. Indeed, preliminary simulations show a clear reduction of the observed non-additivity when the barrier is wider in comparison to the interaction range. Note, however, that the interaction range in our setup is controlled directly via the applied external field, rather then by the inherent properties of the colloidal particles. Such interactions can be induced in a wide range of colloids or nanoparticles, as long as they are susceptible to polarization by an external (electric or magnetic) field. As a second requirement, the density should be high enough to enable significant ordering of particles. In the confined region between the barriers, the ordering will depend sensitively on the ratio of the barrier spacing $\Delta x$ and the preferred spacing between neighboring layers of particles, as long as $\Delta x$ is small compared to the correlation length in the system. Similar confinement effects have shown to result in oscillatory behavior in forces between plates or spheres immersed in a background of smaller particles \cite{roth2000depletion}. Interestingly, the effect of negative resistance is reminiscent of the interplay between reflecting barriers in quantum-mechanical systems, where interference is known to lead to enhanced transmission for certain barrier spacings, as used in e.g. Fabry-Perot interferometers \cite{vaughan1989fabry}.

The sensitivity of the resistance to the barrier separation and microscopic particle interactions provide a method to tailor and control the flow of particles through complex environments \cite{lindner_schimanskygeier_prl2002, gernert_klapp_pre2015, carusela_rubi_jcp2017, spanner_franosch_prl2016, shrivastav2019anomalous}. Indeed, for colloids driven across disordered energy landscapes \cite{hanes2012colloids, peter2018crossover, stoop2018clogging}, long-range interactions have been shown to dramatically affect clogging behavior \cite{stoop2018clogging}. For more periodic energy landscapes, clever choices of the particle interactions and external fields can lead to complex individual or collective dynamics \cite{tierno2016enhanced,mirzaee2020colloidal, loehr2016topological, tierno2009colloidal}, including e.g. an effective negative particle mobility \cite{ros2005brownian}. 

The mitigation of a flow-resisting barrier by placing another barrier near it might have important implications in the design of microfluidic devices, where clogging can be a major issue \cite{clogging, van2016transition, hidalgo2018flow, marin2018clogging, souzy2020clogging}. Moreover, this strategy may also be effective in aiding the flow of particles through geometric constrictions \cite{constriction_paper, brun2019extreme, Liu_Nagel_Nature}, where particles have to pass through a bottleneck rather than over a potential energy barrier. In this scenario additional geometric obstacles -- typically placed before the bottleneck -- have already been shown to enhance flow \cite{alonso2012bottlenecks}, as applied in e.g. the design of emergency exits \cite{tanimoto2010study, shiwakoti2013enhancing}. Hence, it seems likely that potential energy barriers, e.g. induced by external fields, could accomplish the same thing. The specificity of this approach to relatively long-ranged interactions suggests an opportunity for separating different particle species, or enhanced flow control via external fields modifying the interactions or particle motion.  Further applications include the directed transport of strongly charged dust particles in a plasma \cite{Morfill} and congestion in granulates \cite{granulates}, as well as jammed flow situations of colloids \cite{kanehl_stark_prl2017}.  An interesting question for future research is whether the effective total resistance could be further tuned by using a combination of three, four, or an infinite number of obstacles \cite{siems_nielaba_pre2015, prakash2020rapid} (forming e.g. a ratchet \cite{Haenggi,evstigneev2009interaction}), barriers of different heights or shapes \cite{chupeau2020optimizing}, time-dependent barriers \cite{abbott2019transport}, or active particles \cite{koumakis2014directed}.

We gratefully acknowledge funding from the German Research Foundation (DFG) within project LO 418/19-1. We thank Arjun Yodh, Laura Filion, and Marco Heinen for helpful discussions.

\section{Methods}
\footnotesize

{\bf Experimental setup:} Our experiments are based on repulsive microscopic particles, gravitationally driven through microchannels. We use superparamagnetic
colloidal particles (Dynal M-450, diameter $\sigma =4.50(5)$ $\mu$m, $\rho_{m} = 1500\,\text{kg}\,\text{m}^3$) which are
restricted to two-dimensional in-plane motion due to gravity. The cell consists of two rectangular reservoirs of side length $1$ mm which are connected by multiple channels.
The dimensions of each channel is $2$ mm in length, $30$ $\mu$m in width and $8$ $\mu$m in height. In the channels, U--shaped
step-like barrier structures are implemented along the channel, each of them with width $3$ $\mu$m
and height $500$ nm near the channel walls and $250$ nm in the middle of the channel.

The applied magnetic field $\mathbf{B}_\text{ext}$ induces a dipole-dipole repulsion among the colloidal particles, and 
the strength of the dipole-dipole interaction can be tuned by changing the magnitude of the magnetic field.
The repulsive in-plane interaction potential $V(r)$ is \cite{Zahn1997, Microchannels_I}
\begin{align}
V(r) = \begin{cases} \frac{\mu_0 (\chi_\text{eff} \mathbf{B}_\text{ext})^2 }{4\pi r^3}, \quad &\text{for } r \geq \sigma,\\
	                  	\infty, \quad &\text{for } r < \sigma,
	                  \end{cases}
\end{align}
where $\mu_0$ is the vacuum permeability  and $\chi_\text{eff} = 7.88(8) \cdot 10^{-11} \text{Am}^2\text{T}^{-1}$ is
the effective magnetic susceptibility  of the particles. Note that for sufficiently high field strengths, the particles never touch, such that the hard-core component of the interaction potential can be neglected.

By tilting the whole experimental setup, gravity acts as an external driving force, with a strength controlled by the tilt angle and the buoyancy-corrected effective mass of the particles ($m^{*} = 2.385(80) \cdot 10^{-14}$ kg). Using video microscopy, we measure the total particle flux through channels with zero, one, or two barriers as a function of the strength of the driving force.

\medskip

{\bf Channel model}:
The external potential $V_\text{ext}(x, y)$ is composed of a confining channel contribution, $\Vch(y)$, and the barrier potential, $\Vbar(x)$.

The steep repulsive potential forming the channel walls is given by
\begin{align}
 \Vch(y) = V_c \left[ 1 - \frac{1}{2}\erf\left( \frac{y + \frac{L_y}{2}}{\sqrt{2}w} \right) + \frac{1}{2}\erf\left( \frac{y - \frac{L_y}{2}}{\sqrt{2}w} \right) \right],
\end{align}
with channel width $L_y$ and maximum channel potential height $V_c = 1000 k_B T$. The parameter $w = 0.25 a$ sets the softness of the walls. We choose $L_y = 4.65 a$. The channel length $L_x = 25.79 a$ is fixed by the imposed number of particles $N=120$.

A single barrier potential is given by
\begin{align}
    \Vbar(x) = \begin{cases}
                V_0 \left[ 1 - \Big(\frac{\displaystyle{x - x_1}}{\displaystyle{a / 2}}\Big)^2 \right],  &\text{for } |x - x_1| < a / 2\\
                0,   &\text{otherwise}
            \end{cases},
\end{align}
where $x_1$ is the position of the barrier. The double barrier potential is simply the superposition of two non-overlapping single barrier potentials at $x_1$ and $x_2$, where $|x_1 - x_2| = \Delta x \geq a$.

\medskip

{\bf Dynamical density functional theory:}
Within the DDFT framework \cite{DDFT_Tarazona,DDFT_Evans}, the number density field $\rho({\bf r},t)$ of the colloidal particles is calculated by solving the differential equation
\begin{align}\label{DDFT}
	\frac{\partial \rho (\mathbf{r},t)}{\partial t} &= D \nabla \left(\rho(\mathbf{r},t) \nabla \frac{\delta \mathcal{F}[\rho(\mathbf{r},t)]}{\delta \rho(\mathbf{r},t)}\right),
\end{align}
where $D=k_\text{B}T/\xi$ is the single particle diffusion constant, $\xi$ the friction coefficient
and $\mathcal{F}[\rho] = \mathcal{F}_\text{id}[\rho] + \mathcal{F}_\text{ext}[\rho] + \mathcal{F}_\text{exc}[\rho]$ is the total Helmholtz free energy functional. This functional incorporates 
the ideal gas contribution
\begin{align}
    \mathcal{F}_\text{id}[\rho] = k_\text{B}T \int \text{d}\mathbf{r} \; \rho(\mathbf{r}) \big( \log(\Lambda^2 \rho(\mathbf{r})) - 1 \big)    
\end{align}
and the external potential term
\begin{align}
    \mathcal{F}_\text{ext}[\rho] = \int \text{d}\mathbf{r} \; \rho(\mathbf{r}) \big( V_\text{ext}(\mathbf{r}) - x F\big),
\end{align}
where $\Lambda$ is the thermal de Broglie wavelength. 
As an approximation for the excess free energy functional $\mathcal{F}_\text{exc}[\rho]$ we chose the Ramakrishnan--Yussouff functional \cite{RY_functional}
\begin{align}
 \mathcal{F}_\text{exc}[\rho] =& \,\, \mathcal{F}_\text{exc}^\text{ref} (\rho_0) -  \nonumber\\
 & \frac{k_\text{B}T}{2} \int  \!\! \text{d}\mathbf{r} \int \!\! \text{d}\mathbf{r'} \;\Delta\rho(\mathbf{r})\Delta\rho(\mathbf{r'}) c^{(2)}_0(|\mathbf{r} - \mathbf{r'}|;\rho_0,\Gamma).
\end{align}
Here, $\mathcal{F}_\text{exc}^\text{ref} (\rho_0)$ is the excess free energy of an isotropic and homogeneous reference fluid at density $\rho_0$, $\Delta \rho(\mathbf{r}) = \rho(\mathbf{r}) - \rho_0$ describes the density difference to the reference density, and $c^{(2)}_0(r;\rho_0,\Gamma)$ is a pair (two-point) direct correlation function\cite{HansenMcdonald} that has been calculated via liquid integral theory with Rogers-Young closure \cite{Svens_paper}.

 The DDFT is solved numerically by using finite volume difference methods \cite{FiPy}. In each run, we first compute the equilibrium configuration of the system at a given barrier configuration in absence of a driving force ($F = 0$). Then, we switch on the driving force and let the system evolve towards its steady state.

\medskip

{\bf Analytical prediction for non-interacting particles:}

For non-interacting particles the excess free energy vanishes, i.e. $\mathcal{F}_\text{exc} \equiv 0$, and the DDFT in the steady state can be reduced to a single variable $x$. The general solution for periodic boundary conditions and tilted potential $V(x) = \Vbar(x) - Fx$ is \cite{Risken1996}
\begin{align}
    J = \frac{D\rho_0 L_xL_y (1 - \eL)}{I_{+}I_{-}  - (1-\eL)\int_0^{L_x}\d x e^{-\beta V(x)} \int_0^{x}\d {x'} e^{-\beta V(x')}},
\end{align}
with $I_{\pm} = \int_0^{L_x}\d x e^{\pm\beta V(x)}$.

For single and double barrier potentials we can find an analytic expression for $J$ and therefore express the ratio of resistances as
\begin{align}
    \frac{R_2}{R_1} = 1 - K \left(e^{-F(\Delta x - a)} + e^{-F(L_x-\Delta x-a)}\right).
\end{align}
Here, the value $K=\frac{P}{Q}$ does not dependent on $\Delta x$ and is determined by the expressions
\begin{align}
    P &= \beta^2 F^2 (A_1(1-\ea) - A_2) - (1-\ea)^2,\\
    Q &= \beta^2 F^2 (A_1(1-\eaL) + A_3 - A_4) \nonumber\\
    &\quad - \beta F a (1-\eL) - (1-\eaL)(1-\ea),
\end{align}
with
\begin{align}
    \gamma &= \frac{a}{4}\sqrt{\frac{\pi}{\beta V_0}},\\
    \zeta_{\pm} &= \sqrt{\beta V_0}\left(\frac{Fa}{4\beta V_0} \pm 1\right),\\
    \derf &= \erf(\zeta_{+}) - \erf(\zeta_{-}),\\
    \derfi &= \erfi(\zeta_{+}) - \erfi(\zeta_{-}),\\
    A_1 &= \gamma\left(e^{-\zeta^2_{+}} \derfi + e^{\zeta^2_{-}} \derf\right),\\
    A_2 &= \gamma^2\ea\derf\derfi,\\
    A_3 &= \gamma^2\derfi\left(\erf(\zeta_{+}) - \erf(\zeta_{-})\eL\right),\\
    A_4 &= \frac{a^2(1-\eL)}{8\beta V_0}\left(\zeta_{+}^2 \Phi(\zeta_{+}^2) - \zeta_{-}^2 \Phi(\zeta_{-}^2)\right),
\end{align}
where $\erfi(x)$ is the imaginary error function, and $\Phi(x) = \hyp{x}$ is the generalized hypergeometric function.
\medskip

{\bf Brownian dynamics simulations:} 
In addition to DDFT, we perform overdamped Brownian Dynamics simulations of particles. Here, we numerically solve the equations of motion for the particles, given by
\begin{equation}
 \dot{\mathbf{r}}_i = \frac{-\nabla_i {V}_\mathrm{tot}}{\xi} + \frac{F \hat{\mathbf{x}}}{\xi}+ \sqrt{2 D} \mathbf{R}(t),
\end{equation}
where $\mathbf{r}_i$ is the position of particle $i$, and $V_\mathrm{tot}$ is the total potential energy of the system, including particle-particle, particle-wall, and particle-barrier interactions. Finally, $\mathbf{R}(t)$ is a delta-correlated random variable with zero mean and unit variance. The dipolar interactions were truncated and shifted at a distance of $5 a$.

Most simulations were performed using $N=120$ particles, in a channel with periodic boundary conditions along the $x$-axis. We have confirmed that our results are qualitatively the same for larger systems of $N=600$ particles.

\begin{acknowledgement}
We gratefully acknowledge funding from the German Research Foundation (DFG) within project LO 418/19-1. We thank Arjun Yodh, Laura Filion, and Marco Heinen for helpful discussions.
\end{acknowledgement}




\providecommand{\latin}[1]{#1}
\makeatletter
\providecommand{\doi}
  {\begingroup\let\do\@makeother\dospecials
  \catcode`\{=1 \catcode`\}=2 \doi@aux}
\providecommand{\doi@aux}[1]{\endgroup\texttt{#1}}
\makeatother
\providecommand*\mcitethebibliography{\thebibliography}
\csname @ifundefined\endcsname{endmcitethebibliography}
  {\let\endmcitethebibliography\endthebibliography}{}

\end{document}